\documentclass[aps,pre,preprint,superscriptaddress,amsmath,amssymb,floatfix]{revtex4}
\usepackage{bigints}
\usepackage{times}
\usepackage{natbib}
\usepackage{graphicx}
\usepackage{caption}
\usepackage{graphicx,amsmath,amssymb,amstext,amsfonts,color}
\usepackage{hyperref}
\usepackage{movie15}

\begin{document}
\def\D{\Delta}
\def\d{\delta}
\def\r{\rho}
\def\p{\pi}
\def\a{\alpha}
\def\g{\gamma}
\def\ra{\rightarrow}
\def\s{\sigma}
\def\b{\beta}
\def\e{\epsilon}
\def\G{\Gamma}
\def\om{\omega}
\def\l{\lambda}
\def\f{\phi}
\def\w{\psi}
\def\m{\mu}
\def\t{\tau}
\def\c{\chi}

	\title {Competition-driven evolution of organismal complexity }
	
\author{Iaroslav Ispolatov}
\email{jaros007@gmail.com}
\affiliation{
Departamento de Fisica, Universidad de Santiago de Chile,
Casilla 302, Correo 2, Santiago, Chile}

\author{Evgeniia Alekseeva} 
\affiliation{Skolkovo Institute of Science and Technology, Moscow, Russia}

\author{Michael Doebeli}
\email{doebeli@zoology.ubc.ca}
\affiliation{Department of Zoology and Department of
  Mathematics,  University of British Columbia, 6270 University Boulevard, Vancouver B.C. Canada, V6T 1Z4}

	\begin{abstract}
Non-uniform rates of morphological evolution and evolutionary increases in organismal complexity, captured in metaphors like ``adaptive zones'', ``punctuated equilibrium'' and  ``blunderbuss patterns'',  require more elaborate explanations than a simple gradual accumulation of mutations. Here we argue that non-uniform evolutionary  increases in phenotypic complexity can be caused by a threshold-like response to growing ecological pressures resulting from evolutionary diversification at a given level of complexity. Acquisition of a new phenotypic feature allows an evolving species to escape this pressure but can typically be expected to carry significant physiological costs. Therefore, the ecological pressure should exceed a certain level to make such an acquisition evolutionarily successful. We present a detailed quantitative description of this process using a microevolutionary competition model as an example. The model exhibits sequential increases in phenotypic complexity driven by diversification at existing levels of complexity and the resulting increase in competitive pressure, which can push an evolving species over the barrier of physiological costs of new phenotypic features. 
	\end{abstract}
	
	\maketitle
	
\section{Introduction}
Evolution has created  an incredible amount of biological diversity. Clearly, some of this diversity evolved as a consequence of adaptation to various types of resource and habitat conditions. Thus, throughout the history of life on Earth, and despite their relative simplicity, bacteria have remained the most diverse domain, inhabiting even the most inhospitable corners of our planet. However, adaptation to new types of resources or environmental conditions is not the only way of diversification. Competing with and preying on each other, some organisms became more complex and, using their new phenotypic traits, increased their competitiveness and efficiency of resource use. 
Here we use the term ``complexity'' in a rather imprecise and intuitive way, which implies that one organism is more complex than another if it has more diverse and better regulated metabolic functions, more advanced capabilities to sense stimuli, additional means of locomotion, etc. In this context, a reasonable proxy for the complexity of a multicellular organism would e.g. be the number of different  tissues in its body plan.

Generally speaking, it seems reasonable to think that an increase in complexity would entail energy costs for reproduction and development, but that as a result of increased complexity, organisms can achieve significant ecological and, on larger time scales, evolutionary advantages. Double fertilization in flowering plants, breathing of atmospheric oxygen in amphibians, separation of blood circulation in reptiles, and development of speech in humans are just a few examples of such aromorphosis. \cite{moczek2008origins} 

With recent improvements in sequencing technology, knowledge
accumulates about the mechanisms of diversification within many
different groups, representing different levels of
complexity. However, still little seems to be known about the
evolution of  significant innovations and the associated evolutionary
changes in complexity \cite{mayr1960emergence,
  wagner2010evolutionary}. What seems clear is that the evolution of
complexity takes time: about 3 billion years passed  between the
emergence of life and the appearance of the first multicellular
organisms, the first land-based species only appeared about 450 millions years ago, and Archaeopteryx, the first known flying creature, appeared a mere 150 million years ago. 

It has been argued that significant increases in organismal complexity
that open a new ecological niches should often be followed by rapid
diversification and subsequent saturation of diversity in that niche
(e.g \cite{uyeda2011million, schluter2000,
  doebeli_ispolatov2017}). The rates of evolutionary change and of
speciation is expected to be high when the niche is newly formed and
almost empty, and become lower as the niche gets filled through
diversification \cite{schluter2000, doebeli_ispolatov2017}). Such
patterns are sometimes referred to as ``punctuated equilibrium'' or
``blunderbuss pattern'' \cite{gould_eldredge1977,uyeda2011million}, indicating that a uniform accumulation of mutations could not account for intermittent evolutionary bursts. However, what sets the pace of such repeating bursts remains unclear. A number of times in the Earth history, the formation of new niches was caused by geological or cosmic catastrophes, yet it also seems highly likely that such patterns can be caused by intrinsic dynamics of the evolving biosphere itself.

 To an even greater extent than the evolution in general, the evolutionary increases in
complexity
is a relatively slow endeavour that depends on two
processes: One being more probabilistic and mutation-dependent, while
the other being more deterministic and selection-dependent.

1. Evolution of the phenotype is impossible without
mutagenesis. Mutations occur with a  certain probabilities at random
locations in the genome. Single nucleotide substitutions are the most
likely mutation events,  and most of the genetic differences within a
population result from such polymorphisms. Other mutational events,
such as gene duplications or chromosomal rearrangements, are quite
rare, but genomic analysis of closely related  species shows that such
mutations seem to be very important in the history of
aromorphosis. The transition to a new level of phenotypic complexity
requires accumulation of a set of particular mutations that usually
have to come in a certain sequence to be useful and get fixed. Within
this paradigm, the
role of selection is considered to be weak and limited to the ``removal of bad
mutations'' \cite{lynch2007frailty}. Those sequences of specific mutations can have very low probabilities and thus require long waiting times to occur.  

2. Evolution of complexity will depend  on the accumulation of a number of mutations with successive fixation.  Since mutations often carry certain physiological cost, their fixation is conditional on the competitive advantage they provide. The competitive advantage becomes more valuable when ecological interactions,  such as competition for resources or predation, intensify. When the level of diversity at an existing level of phenotypic complexity saturates or becomes sufficiently high, and the competition or predation pressure increases beyond a certain threshold, a mutation or a sequence of mutations that increase complexity but come at a cost can become fixed. Propelling their bearers to a higher level of complexity, such mutations enable them to explore new resources and thus reduce competition, or develop new means to escape predation.

Here we focus on the  role of ecological processes for aromorphosis. Based on logistic competition models, we quantitatively explore how the intensity of ecological interactions  can drive increased organismal complexity despite physiological costs. 
Increases in organismal complexity are described as new dimensions in phenotype space that are acquired during evolution, while the level of diversity is reflected by the number of distinct species.  The physiological cost of adding a new phenotypic dimension is implemented as a reduction in birth rates, while the competitive advantage gained as a result of such addition is modeled as an increase in the environmental carrying capacity.
We show that when the physiological costs of adding a new phenotypic capability are comparable to the benefits that a carrier of the corresponding fully developed capability can gain, then the initial increase in complexity, i.e., the initial gradual acquisition of the new phenotypic dimension or capability, can indeed be driven by ecological interactions.

\section {The model}
 To model the evolution of complexity due to ecological interactions, we study a general class of models for
 frequency-dependent competition \cite{doebeli_ispolatov2014,
   ispolatov_etal2015, ispolatov_etal2015a,
   doebeli_ispolatov2017}, in which ecological interactions are
 defined by continuous $d$-dimensional phenotypes, where $d\geq1$. For example,
 one can imagine that dimensions in phenotypes $\mathbf x$ of
 individuals are given by the efficiencies of several metabolic
 pathways, or various morphological characteristics.  An acquisition
 of a new phenotypic capability is viewed as an expansion into a new phenotypic dimension that represents the new function. 

Competitive ecological interactions that define the logistic model
 are determined by a competition kernel $\alpha(\mathbf{x}, \mathbf{y})$ and a carrying capacity
 $K(\mathbf{x})$, where $\mathbf x,\mathbf y$ 
are the phenotypes of competing individuals. 
The competition kernel $\alpha(\mathbf{x}, \mathbf{y})$
 measures the  competitive impact that an individual of phenotype
 $\mathbf x$ has on an individual of phenotype $\mathbf y$, and in the
 sequel we always assume that   $\alpha(\mathbf{x}, \mathbf{x})=1$ for
 all $\mathbf x$. To take into account the physiological cost of
 maintenance of new phenotypes, we extend the model considered in
 \cite{doebeli_ispolatov2014, ispolatov_etal2015, 
   doebeli_ispolatov2017} by adding a phenotype-dependent birth rate
 $\b (\mathbf x)$. Then the logistic ecological dynamics 
for individuals with phenotype $\mathbf x $ in the environment with
phenotypes $\mathbf{y_r}$ is completely determined by
the birth rate $\b(\mathbf x)$ and the death rate
\begin{align}
\label{birthdeath}
\frac{\sum_{r} \a(\mathbf{y_r},\mathbf{x})}{K(\mathbf x)}.
\end{align}

To make our arguments more clear and simplify the analysis, we apply
the standard adaptive dynamics approach and calculate the
invasion fitness, i.e., the per capita growth rate of a rare mutant with
phenotype $\mathbf y$ in the resident monomorphic population with
phenotype $\mathbf x$, 
 \begin{align}
 \label{fitness}
f(\mathbf{x}, \mathbf{y}) = \beta(\mathbf{y}) - \frac{ \alpha(\mathbf{x}, \mathbf{y}) K(\mathbf{x})}{K(\mathbf{y})}.
\end{align}
 With
the resident population consisting of several clusters $r=1,\dots$ with distinct phenotypes $\mathbf{y_r}$, the adaptive dynamics for a cluster with phenotype $\mathbf x $
is determined by its selection gradient $\mathbf {s(x)}$ with components
 \begin{align}
\label{sg}
s_i(\mathbf{x}) \equiv \left. \frac{\partial f(\mathbf{x}, \mathbf{z})}{\partial z_i}\right|_{ \mathbf{z}=\mathbf{x}} = 
\frac{\partial \b (\mathbf x)}{\partial x_i}
 - \left. \sum_{r} \frac{N_r}{K(\mathbf{x})} \frac{\partial \alpha(\mathbf{y}_r,
   \mathbf{z})}{\partial z_i}\right|_{
   \mathbf{z}=\mathbf{x}} + \sum_{r} \frac{\partial
   K(\mathbf{x})}{\partial
   x_i}\frac{\alpha(\mathbf{y}_r,\mathbf{x})N_r}{K^2(\mathbf{x})},
\end{align}
(see \cite{doebeli2011, doebeli_ispolatov2014, ispolatov_etal2015, doebeli_ispolatov2017} for more details).
Here $N_r$  is the equilibrium population of the
cluster with phenotype $\mathbf{y_r}$, which is given by the
stationary solution of the system of logistic population dynamics
equations,
 \begin{align}
\label{Ni}
\frac{dN_r}{dt} = N_r\left[\b (\mathbf {y_r})  - \frac {\sum_{r'} \alpha(\mathbf{y}_{r'},
 \mathbf{y}_r)N_{r'}}{K(\mathbf{y}_r)}\right].
 \end{align}
The selection gradients define a system of differential equations in phenotype space $\mathbb{R}^d$, 
 \begin{align}
 \label{AD1}
 \frac{d\mathbf{x}_r}{dt} = N_r\mathbf{s}_r(\mathbf{x}_r).
\end{align} 
For simplicity and generality, here we assumed that  the mutational variance-covariance matrix for each cluster, which
reflects peculiarities of genotype-phenotype mapping, is diagonal with
elements equal to the population size of the corresponding cluster. This corresponds to the assumption that mutations occur independently in all phenotypic directions, with equal average size and at equal per capita rates. 
More details on the derivation of the adaptive dynamics
(\ref{AD1}) can be found in a large body of original literature
(e.g. \cite{dieckmann_law1996,geritz_etal1998, diekmann2003,
  leimar2009, doebeli2011}). 

The standard adaptive dynamics   is extended  as in \cite{doebeli_ispolatov2017} to include diversification, which manifests itself as the splitting of clusters.
Each $\t_c\sim 1 $ time units the distances between clusters are
assessed and those which are closer to each other than a threshold 
$\D x \sim 10^{-3}$ are merged. Then a new
  cluster is created by randomly picking an existing cluster, 
  splitting it in half, and separating the two new clusters in a
  random direction in phenotype space by
  a distance $\D x$. This ensures a randomly assigned
   capability for all populations to diversify: if a chosen population is under
   selection to undergo diversification, the split clusters
   will diverge sufficiently so that they will not be merged
   back at the next check. Alternatively, when a cluster splits that is not under diversifying selection, the two halves will not diverge and instead will be merged again at the next check.

The key parts in our model are the definitions of  the birth rate $\b (\mathbf {x})$, the competition
kernel $\alpha (\mathbf {x},\mathbf {y})$, and the
carrying capacity $K(\mathbf{x})$, which reflect the costs and
advantages of acquiring new phenotypes.  An acquisition of
an additional phenotypic feature should result in certain
benefits: we express those benefits as an increase in the carrying capacity, resulting in a reduced death rate. However, it
should also be accompanied by a penalty, reflecting higher physiological costs of maintaining a more
complex organism. Here we incorporate these costs in the birth rate. 
Taking these factors into account, our model works as follows:
\begin{itemize}
\item A cluster is considered lacking the phenotypic dimension $i$
  when the corresponding phenotypic coordinate $x_i$ is close to
  zero. For example, one can think of a certain phenotypic dimension
  as of an ability to metabolize a particular substance, so that the
  corresponding phenotypic coordinate is the rate of this  metabolic
  process. The inability to metabolize a substance means that the rate of
  corresponding process is zero. 
\item The carrying capacity has the form 
\begin{align}
\label{K}
K(\mathbf x)= \exp\left[ - \frac{\sum_{i=1}^{d} (x_i-C)^4}{4}\right]  
\end{align}
with the maximum at $\mathbf x=\mathbf C\sim1$. We consider that a
cluster acquires a particular phenotypic dimension $i$ when $x_i$
moves away from zero and gets
sufficiently close $C$. Doing so, the organism makes full use of its
new phenotypic capability by maximizing  the carrying capacity in that
phenotypic dimension.
\item The initially existing simplest life is represented by a single
  cluster that has only one phenotypic dimension and
  the birth rate is the same for all phenotypes along this dimension.
\item Each transition to a higher phenotypic dimension is associated
  with a cost implemented as a reduction 
  in the birth rate.  We define the birth rate $\b(\mathbf x)$ as 
\begin{align}
\label{beta}
\b(\mathbf x)= \prod_{i=2}^d \left\{\exp\left[ - \frac{
  (x_i)^2}{2\s_{\b}^2}\right](1-b) + b\right\}.  
\end{align}
The first phenotypic dimension does not carry any birth rate penalty,
thus the product in (\ref{beta}) starts with $i=2$. When a new dimension
is acquired, that is, the corresponding coordinate changes from almost
zero to a value much larger than the birthrate penalty width
$\s_{\b}$, the birthrate is multiplied by a factor $b<1$.
\item For simplicity, we use the symmetric Gaussian competition kernel 
\begin{align}
\label{alpha}
\a(\mathbf{y},\mathbf{x})=\exp\left[ - \frac{\sum_{i=1}^{d} (x_i-y_i)^2}{2\s_{\a}^2}\right],  
\end{align}
 in which the competitive effect of $\mathbf{y}$ on $\mathbf{x}$
is  equal to that of $\mathbf{x}$ on $\mathbf{y}$. This form of
 competition kernel also promotes expansion into new phenotypes as the
 multiplicative nature of (\ref{alpha}) ensures that the competition
 gets weaker if any of the distances $|x_i-y_i|$ increases. For instance, this
 happens when a cluster acquires the dimension $i$, so that $x_i\sim
 C$, while the rest of the system does not, so that $y_i \approx 0$. 
\end{itemize} 

\section{Adaptive dynamics of acquisition of a new phenotype}
Let us consider a generic scenario of a competition-driven
expansion into a higher phenotypic dimension.  
For simplicity of visualization, we consider the evolutionary transition 
from
1-dimensional to  2-dimensional phenotype space. However, 
the same arguments apply to any increase in phenotypic
dimension. Two components of the selection gradient (\ref{sg}) for a single population
cluster initially living in the first dimension are
 \begin{align}
\label{sg2}
s_1(x_1,x_2)=- (x_1-C)^3 \exp\left[ - \frac{\sum_{k=1}^{2} (x_k-C)^4}{4}\right] \\
\nonumber
s_2(x_1,x_2) = -\frac{x_2}{\s_{\b}^2} \exp\left[ - \frac{
  x_2^2}{2\s_{\b}^2}\right](1-b) - (x_2-C)^3 \exp\left[ - \frac{\sum_{k=1}^{2} (x_k-C)^4}{4}\right]. 
\end{align}
Selecting sufficiently small width $\s_{\b}$ in the birthrate penalty
term makes the contribution of this term dominant and restricts the evolutionary dynamics of the
single cluster to a narrow strip along the $x_1$ axis, $x_2 \ll C$.
In this case, evolution in the first phenotypic dimension follows the standard
adaptive dynamics scenario, i.e. the single cluster moves to the
center of the carrying capacity $x_1^*=C$ and then, for sufficiently
small $\s_{\a}$, evolutionary branching occurs \cite{geritz_etal1997, dieckmann_doebeli1999} with subsequent diversification into two different phenotypic clusters (with each one still having a 1-dimensional phenotype). After splitting, each cluster experiences an additional
contribution $\tilde{\mathbf{s}}$ to its selection gradient that
is produced by the gradient of the competition kernel and is given by the
second term in the right-hand side of (\ref{sg}).
If the two new clusters have phenotypes $\mathbf x $ and $\mathbf y$ and
the distance between them, $|\mathbf x -\mathbf y|$, is sufficiently
small, the components of this contribution are
\begin{align}
\label{sg3}
\tilde s_i(\mathbf x)= \frac{x_i-y_i}{2\s_{\a}^2}\exp\left[ -
  \frac{\sum_{k=1}^{2} (x_k-y_k)^2}{2\s_{\a}}\right],\\
\nonumber
\tilde s_i(\mathbf y)=-\tilde s_i(\mathbf x),\; i=1,2.
\end{align}
The factor 2 in the denominator appears because the population of each of the recently
separated clusters is approximately half of the carrying capacity. 
Assume that for a typical separation $|\mathbf x -\mathbf y|$, the
addition of (\ref{sg3}) to the second-dimensional component of the
selection gradient tilts the balance and makes the  $s_2$ positive for
one of two new clusters. This cluster will start moving in the positive
$x_2$ direction. At the same time, the $\tilde s_1$ components of (\ref{sg3}) is
pushing the two clusters apart in the $x_1$ dimension (under the assumption that $\s_{\a}$ is
small enough to produce diversification). That, in turn, will reduce the
exponential factor in $\tilde s_2$ (\ref{sg3}), which depends multiplicatively on both $| x_1 - y_1|$ and $| x_2 - y_2|$, and at some point will
turn the component $s_2$ negative, resulting in a failed attempt to increase
phenotypic complexity. Such failure thus occurs when diversity in the existing phenotypes is not high enough and the newly split
clusters have ample space to diverge along the $x_1$ coordinate. 

Consider now a complimentary scenario when, as a result of repeated diversification, 
 \cite{doebeli_ispolatov2017}, the $x_1$
dimension has become
saturated with phenotypic clusters, so that  further diversification in the $x_1$-direction  is impossible. This implies that when a new cluster is formed, there
will be no net repulsive $s_1$ component in the selection gradient acting on it. Any such component from a nearest neighbour will be cancelled by competitive repulsion from
other clusters. However, evolution of a newly formed cluster with  $x_2>0$ in the 
positive $x_2$ direction will not be impeded. Rather,  there is competitive pressure to evolve away from $x_2\approx 0$ due to the sum of competitive contributions to $s_2$ generated from all the clusters present in the $x_1$-direction, since their $y_2$ components are close to zero and thus less than $x_2$ (see (\ref{sg3})). The resulting positive contribution to the selection gradient can exceed the negative part of the selection gradient in the $x_2$-direction  that is caused by the cost in the birth rate. Once the competition from the clusters present in the $x_1$-direction has pushed the $x_2$-component of the new cluster  
sufficiently far away from 0, so that $x_2 \gg \s_{\b}$, the negative
part of the selection gradient coming from the cost in the birth rate 
disappears, and $x_2$ continues to increase further and converges to $C$, driven by both an increase in the carrying capacity and by competition from the other clusters.

The essential mechanism underlying the above scenarios is that once
diversity has saturated in a given dimension, the resulting selection
pressure due to competition can be enough to drive evolution into new
phenotypic dimensions despite the necessity of overcoming a fitness
valley that is due to costs of the initial development of the new
phenotypes. These costs are high enough such that without saturation,
competition simply results in further diversification along the
already existing phenotypic dimension. In principle, this should
result in a two-tiered evolutionary process: first, there is
diversification along existing phenotypic dimensions; once this
diversification has saturated, i.e., once the niches along existing
phenotypes are sufficiently full, competitive pressures from the saturated community facilitate the evolution of new phenotypic dimensions that are ``orthogonal'' to existing ones. Acquisition of this new phenotypic dimension should then be followed by another round of repeated diversification in the newly established phenotype space, eventually leading to saturation, which then in turn can again generate further increases in phenotypic complexity. 
The rather complex interplay between the carrying
capacity, the birth rate and the competition kernel in our models do not allow us to make
these argument 
more precise mathematically, but the numerical examples shown in the
next two sections
indicate that such scenarios can indeed be realized for a range of parameters
$\s_{\b}, b, \s_{\a},$ and $C$. 

\section{Results for the transition from one-dimensional to
  two-dimensional systems} 
 
Here we illustrate the arguments made in the previous section with
several numerical examples.
In Figure~1 we show two types of adaptive dynamics evolutionary
trajectories. When the system is initialized with a single cluster
in the first dimension  (the cluster was put close to the maximum of
carrying capacity), it splits into two
clusters, which diverge but both remain in the first dimension (left panel). In
contrast,  when a system has four clusters, which is
the maximum steady state level of diversity for a one-dimensional
system with given parameters, a newly formed cluster moves into the
second dimension (right panel).
        \begin{figure}
		\centering
		\includegraphics[width=3.in]{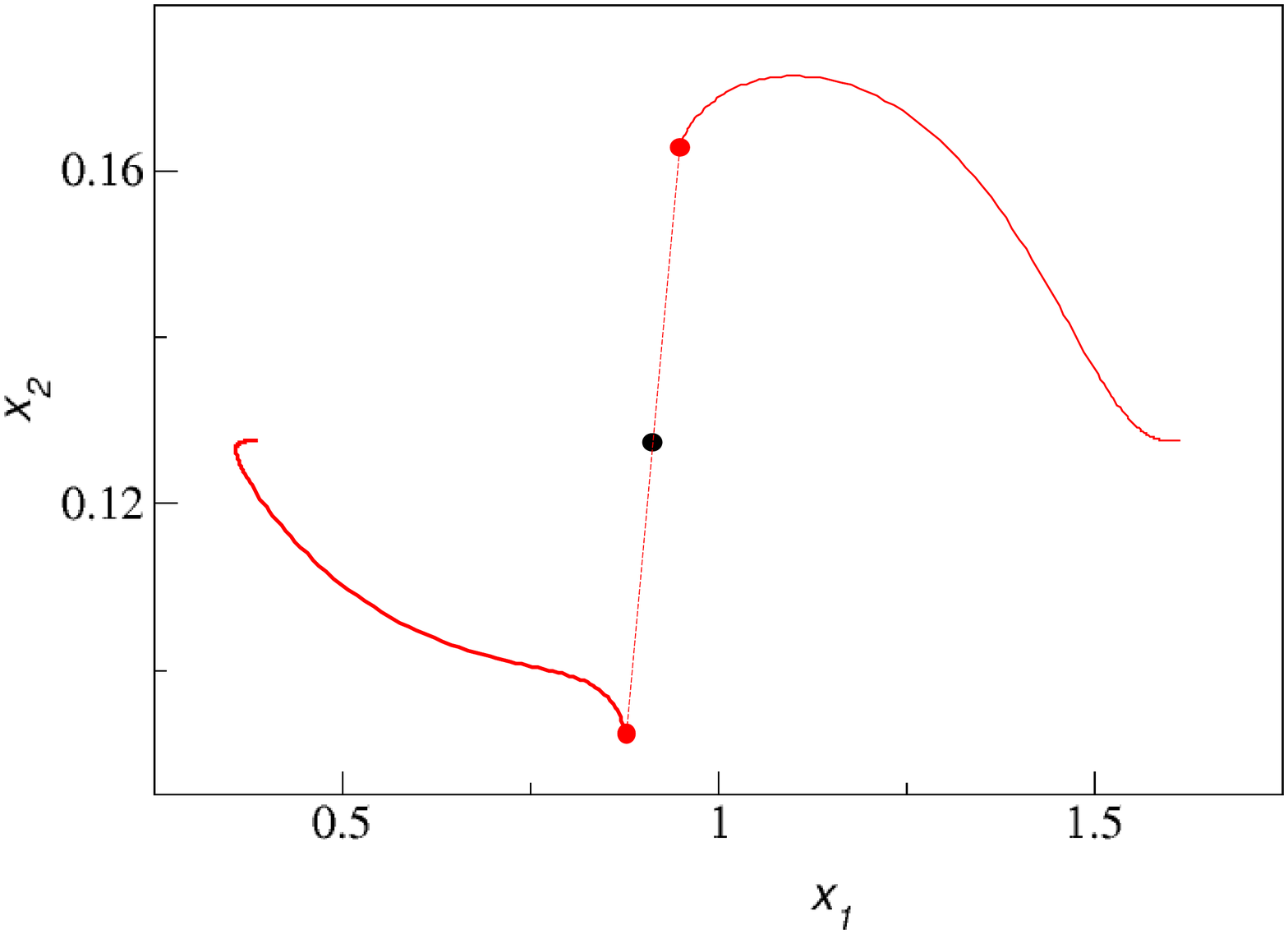}
		\includegraphics[width=3.in]{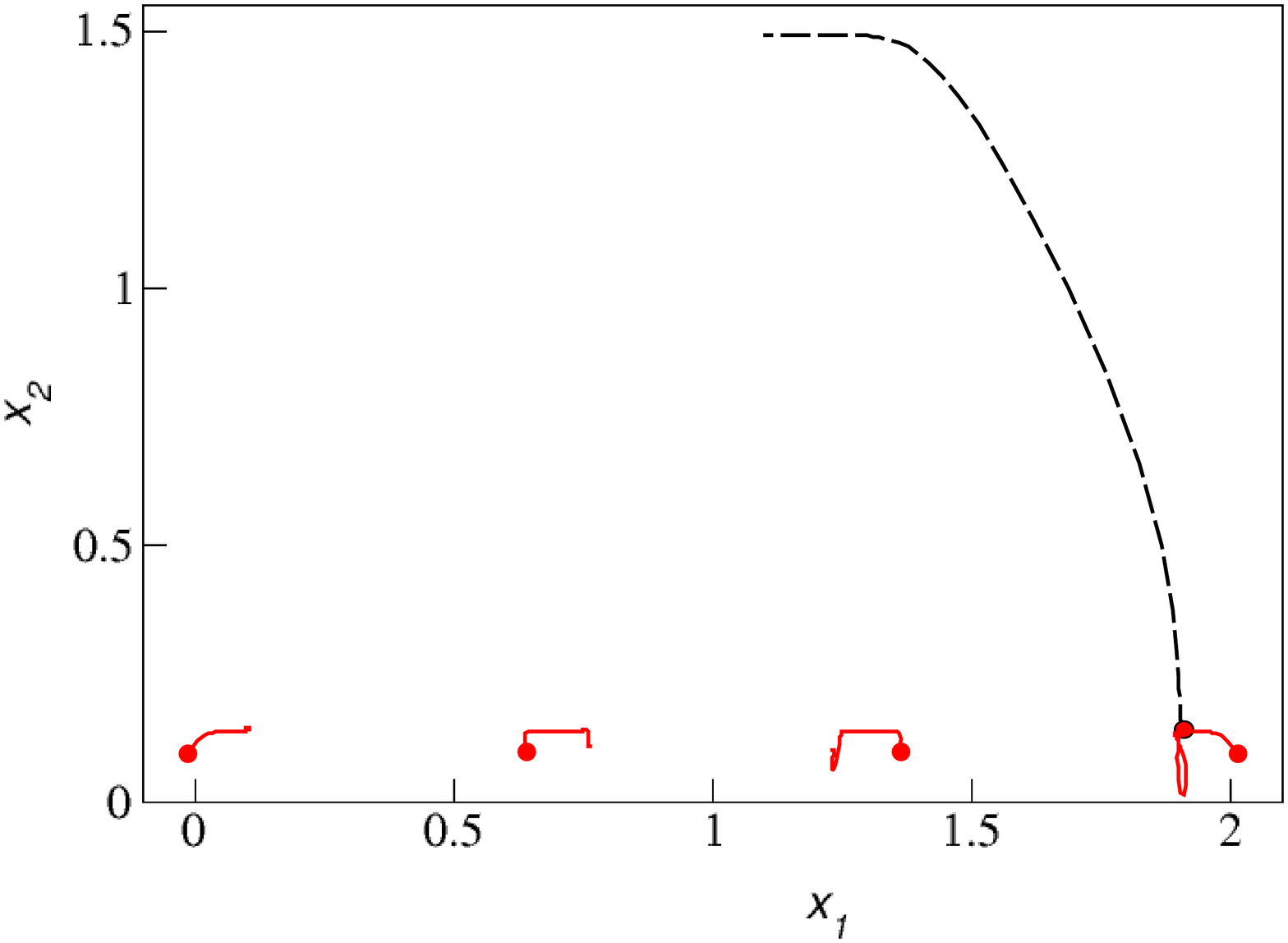}
		\caption{Adaptive dynamics trajectories of
                  diversifying clusters showing failed and successful
                  expansion into the second dimension.
                Left panel: Low diversity scenario of splitting of a single
                one-dimensional cluster (shown by a black circle) in
                two halves (shown by two red circles), which move
                apart but stay in the first dimension (red
                lines). Right panel: Saturated diversification scenario when one of
              four 1-dimensional clusters (whose adaptive dynamics
              trajectories are shown in red) is split in halves and one
              half moves up into the second dimension 
              (its trajectory is shown as black dashed line). Red
              circles show the positions of clusters immediately after the splitting. The
              parameters are $\s_{\b}=0.15$, $b=0.84$, $C=1$, and $\s_{\a}=0.5$. }
		\label{fig1a}
	\end{figure}

A more complete  adaptive dynamics scenario of evolutionary expansion from one
to two phenotypic dimensions is shown in the video in
Fig.~\ref{fig2}, in which a system initialized with a single cluster in the first
dimension first diversifies to saturation in that dimension, and only then expands into
the second dimension. Note that after the first expansion into the second dimension, diversification continues until the available 2-dimensional phenotype space saturates with diversified phenotypic clusters.
       \begin{figure}
		\centering
		\includemovie[poster]{15cm}{15cm}{figure2.avi}
		\caption{Video of adaptive dynamics of diversification
                  showing that an expansion into the second dimension
                  occurs only once the diversity in the first
                  dimension (horizontal strip at the centre of the frame)
                  becomes saturated with 4 distinct clusters. 
                  Parameters are $\s_{\b}=0.15$, $b=0.85$, $C=1$ and $\s_{\a}=0.5$.}
		\label{fig2}
	\end{figure}

To show robustness of these scenarios, we also performed 
individual-based (Fig.~\ref{fig3}) and partial differential equation (Fig.~\ref{fig4})
simulations of the logistic model defined by
(\ref{K},\ref{beta},\ref{alpha}) (see e.g. \cite{ispolatov_etal2015a, doebeli_ispolatov2017}).
Both types of simulations exhibit
the same type of evolutionary dynamics as seen in the adaptive dynamics version:
First, the diversity in the existing dimension becomes saturated,
thus maximizing the competitive pressure, and only then  an expansion into 
the new dimension occurs. After this initial foray,  diversification in
 two-dimensional space continues until saturation. 
       \begin{figure}
		\centering
		\includemovie[poster]{15cm}{15cm}{figure3.avi}
		\caption{Video of an individual-based simulation of the logistic equation defined by the
                  birth rate (\ref{beta}), carrying capacity (\ref{K})
                  and competition kernel (\ref{alpha}). Mutations at
                  birth were implemented as a random offset of the
                  phenotype of the offspring from the ancestral one
                  by $\sim 10^{-2}$, see  \cite{ispolatov_etal2015a,
                    doebeli_ispolatov2017} for more details. It
                  shows that an expansion into the second dimension
                  occurs only when the diversity in the first
                  dimension (a horizontal strip at the bottom of the frame)
                  reaches 4 distinct clusters. Diversification then
                  continues until the two-dimensional space becomes
                  filled with clusters. 
                  The parameters are $\s_{\b}=0.15$, $b=0.75$, $C=1$ and $\s_{\a}=0.5$. }
		\label{fig3}
	\end{figure}	
       \begin{figure}
		\centering
                \includemovie[poster]{15cm}{15cm}{figure4.avi}
		\caption{Video of a partial differential equation 
                  simulation of the logistic equation defined by the
                  birth rate (\ref{beta}), carrying capacity (\ref{K})
                  and competition kernel (\ref{alpha}). A small
                  diffusion term $D \nabla^2 \mathbf x$ with $D=10^{-6}$ is added
                  to mimic mutations. The simulation  
                  illustrates that expansion into the second dimension
                  occurs only when the diversity in the first
                  dimension (a horizontal strip at the bottom of the
                  frame)
                  becomes saturated with an almost continuous 
                  phenotype distribution concentrated around three
                  clusters. 
                  The parameters are $\s_{\b}=0.15$, $b=0.35$, $C=1$ and $\s_{\a}=0.5$. }
		\label{fig4}
	\end{figure}	

\section{Evolutionary expansion into higher dimensions }

\begin{figure}
		\centering
		\includegraphics[width=7in]{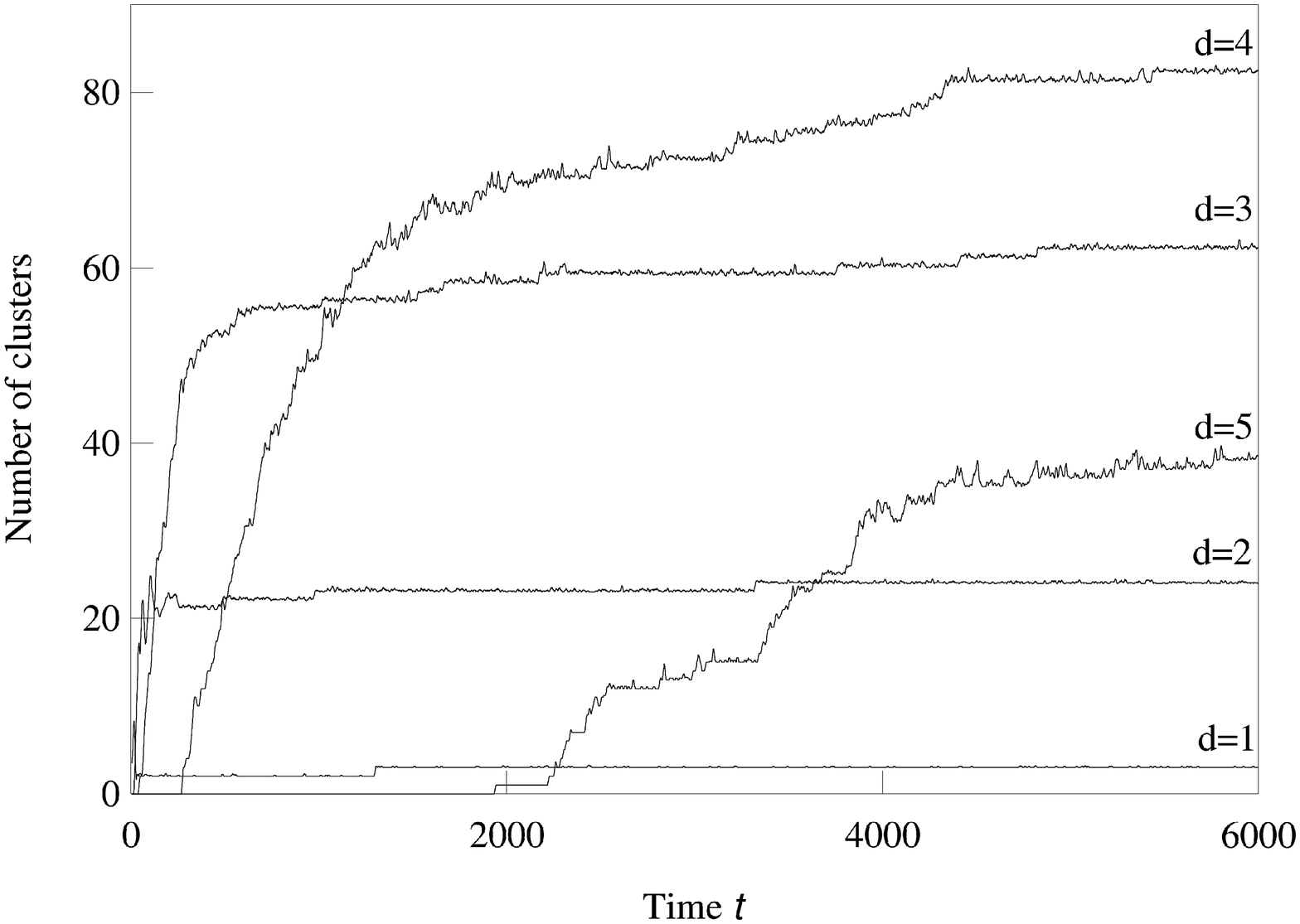}
		\caption{ Adaptive dynamics simulation showing
                  sequential competition-driven expansion
                  into higher phenotypic dimensions. 
                 The numbers of clusters in 1,2,3,4, and 5-dimensional
                  phenotype spaces are
                  shown as a function of time. Here and
                  in the following, a cluster is considered belonging to the
                  dimension $k$ if $x_k> 2 \s_{\b}$ and the running
                   average over 10 time units for the  number of
                  clusters is shown. The  
                  parameters used in the logistic model were
                  $\s_{\b}=0.15$, $b=0.883$, $C=1$, and $\s_{\a}=0.5$.}
		\label{fig5}
	\end{figure}

\begin{figure}
		\centering
		\includegraphics[width=7in]{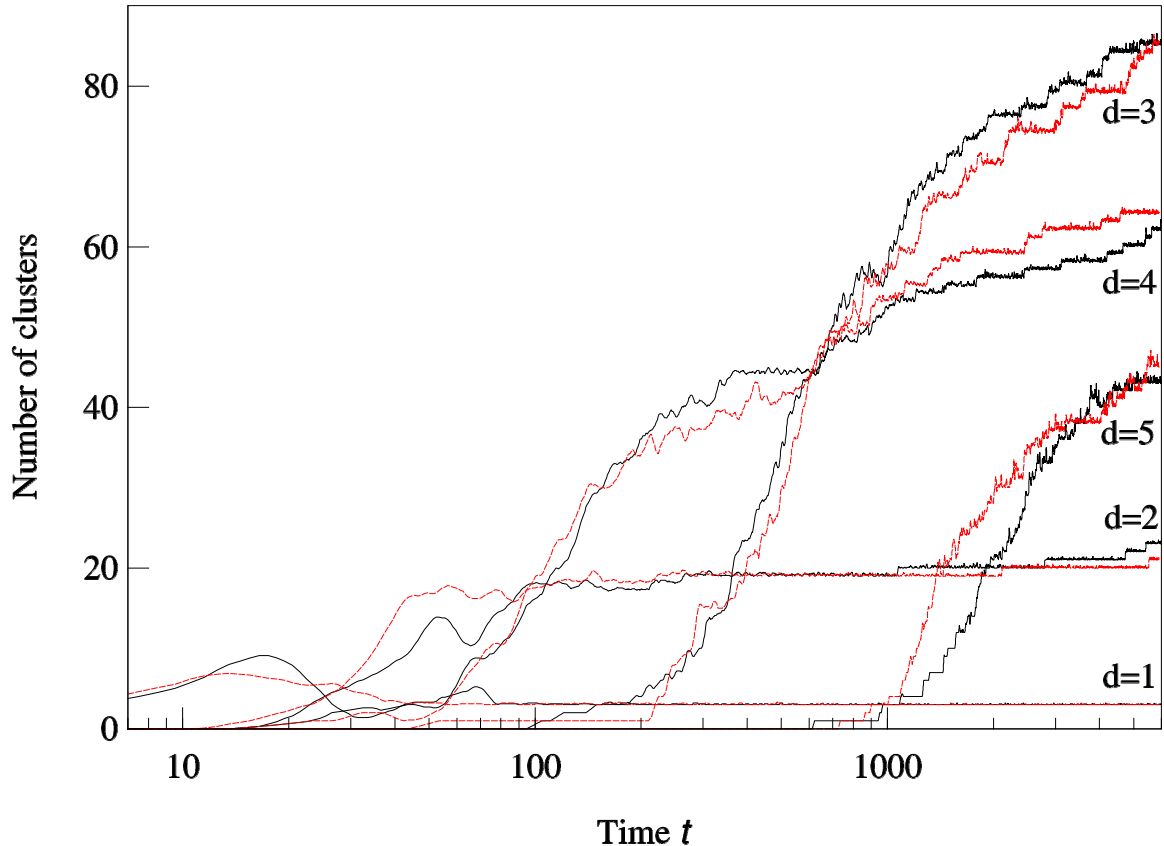}
		\caption{ Two runs of adaptive dynamics simulations
                  showing statistical reproducibility of competition-driven expansion
                  into higher phenotypic dimensions.
                  The numbers of clusters in 1,2,3,4, and 5-dimensional
                  phenotype spaces are shown as a function of time
                  for two distinct runs in solid black and dashed red lines. The
                  parameters used in the logistic model were
                  $\s_{\b}=0.17$, $b=0.9$, $C=0.9$, and $\s_{\a}=0.5$. }
		\label{fig6}
	\end{figure}

\begin{figure}
		\centering
		\includegraphics[width=7in]{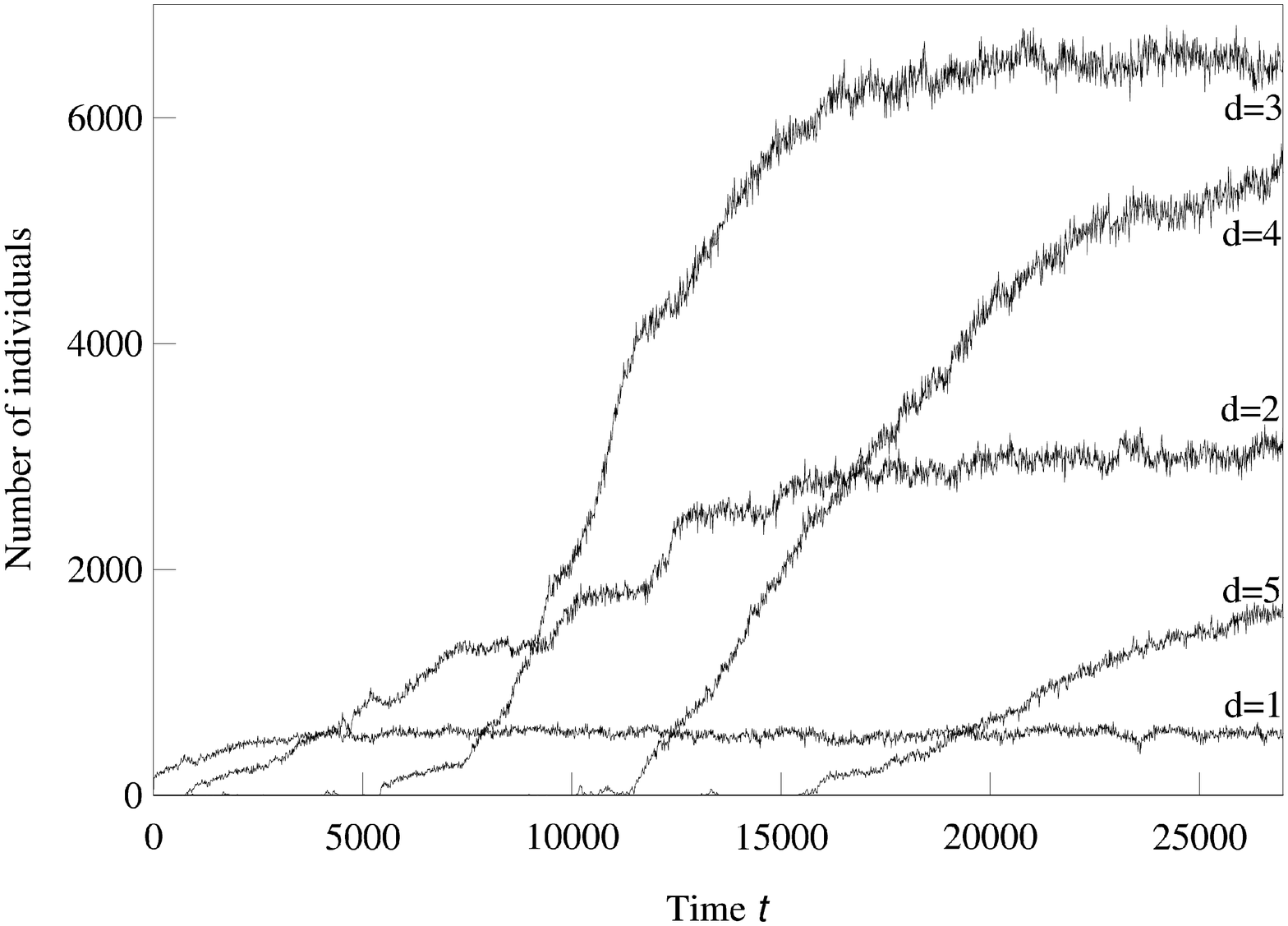}
		\caption{  Individual-based simulation showing
                  sequential competition-driven expansion
                  into higher phenotypic dimensions.  The numbers of individuals 
                  in 1,2,3,4, and 5-dimensional
                  phenotype spaces are shown as a function of time. The
                  parameters used in the logistic model were
                  $\s_{\b}=0.15$, $b=0.75$, $C=1$, and
                  $\s_{\a}=0.5$. The maximum of the carrying capacity $K_0=500$. }
		\label{fig7}
	\end{figure}

We now consider systems in phenotype spaces of dimension larger than two. In our examples we set the highest
dimension equal to five. We expect the
diversification and acquisition of new dimensions to occur as a
sequence of elementary steps similar to the ones presented above for the transition from one to
two dimensions. The number of possible diversification
options naturally increases due to combinatorics: a
population having a one-dimensional phenotype can acquire a two-dimensional
phenotype in four possible ways, expanding into dimension 2, 3,
4, or 5. 
Likewise, there are different ways to expand from
2-dimensional to 3-phenotypes and from 3-dimensional to 4-dimensional phenotypes. The ``elementary expansion events'' corresponding to these various possibilities do not
necessarily happen synchronously due to the stochasticity that is intrinsically
present in our adaptive dynamics diversification procedure
(and even more so in the individual-based simulations). Hence we do not
expect the results to stay precisely as clean as in the case of
expansion from 1 into just 2 dimensions. For example, expansion into
three dimensional space may happen before all two-dimensional
subspaces are completely filled with clusters.  Nevertheless, our results
shown in Figs.~\ref{fig5},\ref{fig6},\ref{fig7} confirm that the general trend remains  
the same: the expansion into a new dimension occurs only when a sufficient level of
diversification and competition pressure is achieved in the existing dimensions.  Figs.~\ref{fig6} illustrates the statistical reproducibility of the scenario of expansion into new dimensions, showing results of two distinct adaptive dynamics runs. In Figs.~\ref{fig7} we show the results of individual-based simulations, which exhibit a sequence of expansion events into higher dimensions that is almost identical to that seen in adaptive dynamics. 

\begin{figure}
		\centering
		\includegraphics[width=7in]{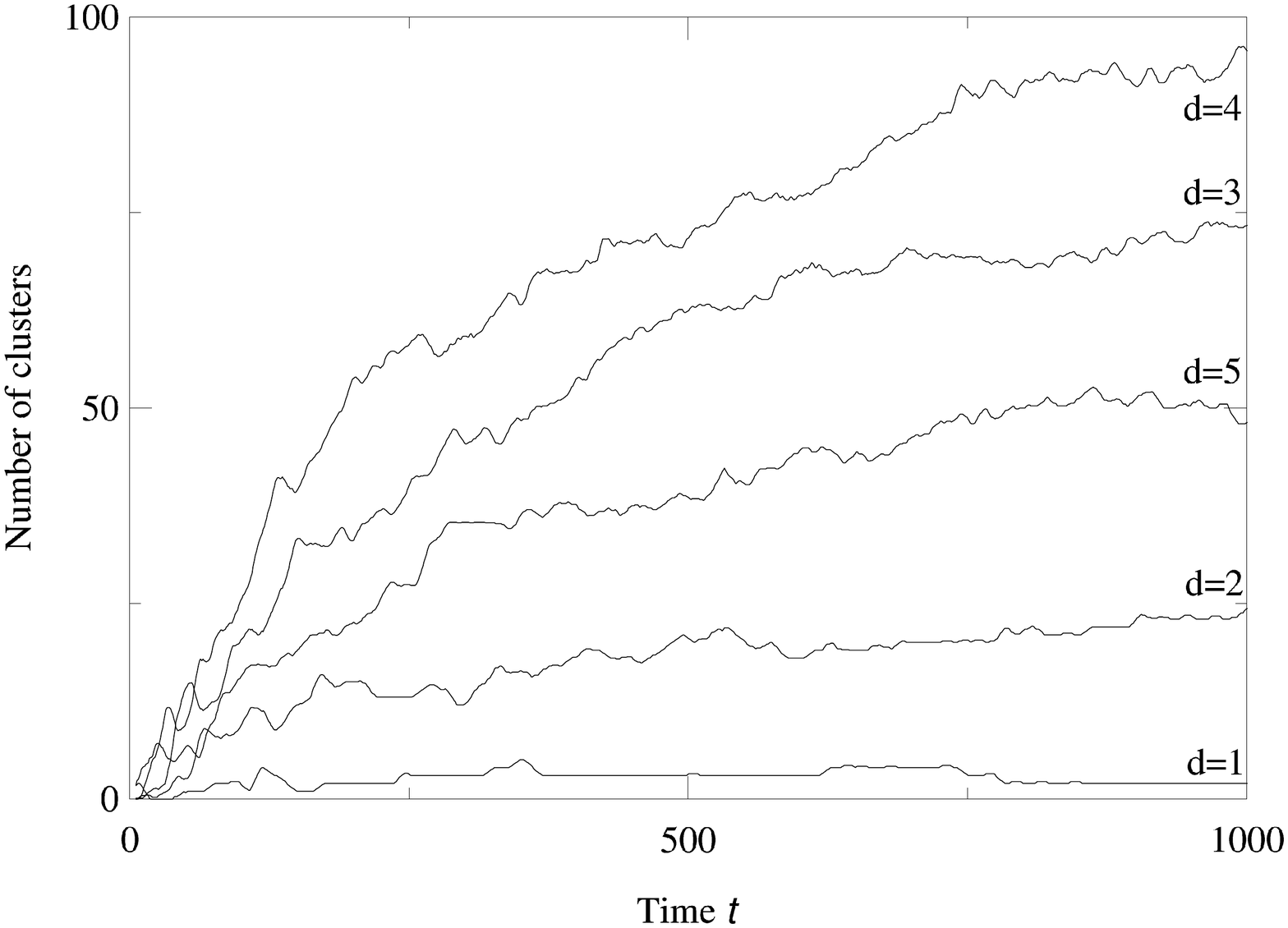}
		\caption{ Adaptive dynamics simulation showing
                  that evolutionary expansion occurs simultaneously
                  into all higher dimensions when there is no cost in
                  the birth rate for increased phenotypic complexity.
                 The numbers of clusters in 1,2,3,4, and 5-dimensional
                  phenotype spaces are
                  shown as a function of time.
                  The
                  parameters used in the logistic model were
                  $\s_{\b}=0.15$, $b=0$, $C=1$, and $\s_{\a}=0.5$.}
		\label{fig_no}
	\end{figure}

To illustrate that the sequential nature of competition-driven evolutionary expansion
into new phenotypic dimensions is indeed conditional on the presence of costs of increased complexity, Fig.~\ref{fig_no} shows an adaptive dynamics simulation where the cost in the birth rates was set to 0. In this case, expansion occurs simultaneously and instantaneously into  all possible phenotypic dimensions, essentially because phenotype expansion only has benefits (both in terms of evading competition and in terms of increasing the carrying capacity). In particular, without costs, the sequential ``blunderbass pattern'' \cite{uyeda2011million} is lost.

\begin{figure}
		\centering
		\includegraphics[width=7in]{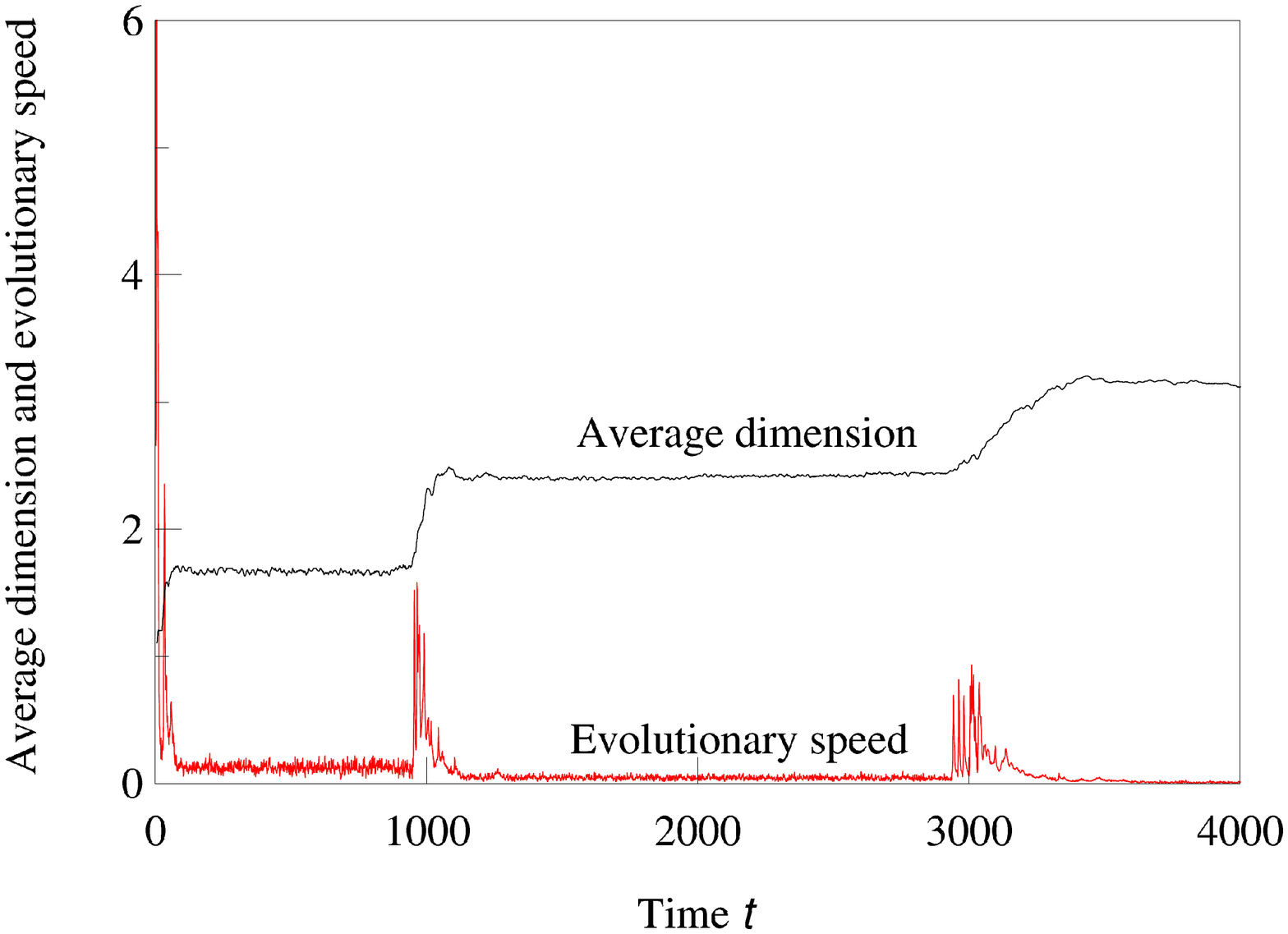}
		\caption{ Correlation between bouts of expansion
                  into new phenotypic dimensions and intermittent increases in the average
                  evolutionary speed. The average dimension is
                  computed as the sum of dimensions of all present
                  clusters divided by the number of clusters. The
                  average evolutionary speed is the population-weighted
                  average of the absolute values of the evolutionary speed (measured by the selection gradients given by adaptive dynamics) of
                  all coexisting clusters. (For purposes of illustration, the speed  
                  is multiplied by 100.). The
                  parameters used in the logistic model were
                  $\s_{\b}=0.16$, $b=0.8$, $C=0.9$, and
                  $\s_{\a}=0.5$.}
		\label{fig8}
	\end{figure}

An interesting consequence of sequential increases in phenotypic complexity as shown in Figs.~\ref{fig5},\ref{fig6} is that
each evolutionary expansion into a new phenotypic
dimension opens up a new and initially empty ecological niche. As predicted in \cite{doebeli_ispolatov2017}, these expansions are not only followed by new bouts of diversification, but they also generate an initial increase in the rate of evolution, which subsequently decreases as diversity in the new niche reaches saturation. Thus, sequential increases in complexity lead to intermittent bursts of evolutionary speed, as illustrated in Fig.~\ref{fig8}.

\section{Discussion}

It seems reasonable and intuitive that evolutionary transitions from simpler to more complex
organisms, capable of accessing novel resources or having otherwise novel ecological properties, 
should go through intermediate phases in which the benefits of
novel phenotypes are not fully available, but the cost of developing those phenotypes is already manifest. Cleary, when costs are
low, transitions can happen fast, with the rate mostly limited by the speed
of accumulation of the necessary mutations. Conversely, when the costs are
very high and cannot be compensated by even the fully accessed
benefits, the transition may never happen. Here we explored intermediate and arguably more intriguing scenarios, in which costs of increased complexity are high enough
to prevent ``trivial'' transitions due beneficial mutations only, yet once
the new phenotype is fully developed, the benefits exceed the costs. 
We have shown that in such an intermediate case, resource competition can make the difference
between success and failure of an evolutionary expansion into a new phenotypic dimension: For
low diversity at the existing level of organismic complexity, competitive pressure is weak, and the expansion
does not occur, while for high diversity in the existing
phenotypic dimensions, competitive pressure becomes strong enough for mutants with higher phenotypic complexity to overcome the costs of increasing complexity. 


This intermediate scenario corresponds to a bistability in the invasion
fitness landscape with two types of maxima: one type (comprising one or many maxima depending on the amount of diversity)
that completely lacks  the new phenotypic 
dimension, while the other type corresponds to the fully developed phenotype in the new dimension. These two types of maxima are separated by a
low-fitness area characterized by incomplete benefits but substantial costs. When
the level of diversification at the existing level of complexity is
low, the corresponding maximum or maxima are high, and the low-fitness
barrier is impassable evolutionarily. But when the level of diversity
in the existing dimension becomes saturated, the invasion fitness maxima in
that dimension flatten and the barrier to the higher dimension
disappears due to competitive pressure from the species living in the lower dimensional phenotype space. Thus, competition enables the crossing into a higher dimensions through an area that is a fitness valley in the absence of competition. 

This work is a continuation of our studies of evolutionary dynamics and speciation in
multidimensional phenotype spaces. We have previously shown that diversification is more likely with high-dimensional phenotypes 
\cite{doebeli_ispolatov2010}, that the evolutionary dynamics even of 
single populations tends to be complicated and  possibly chaotic
\cite{doebeli_ispolatov2014}, that with complex evolutionary dynamics diversification can occur even if the system does not converge to an evolutionary branching point
\cite{ispolatov_etal2015a}, and that for evolution in given phenotype
space, diversification changes a fast evolving community with few
species into a saturated multi-species community whose component
species are evolving only very slowly \cite{doebeli_ispolatov2017}.
Here we have shown that such an evolutionary
standstill may be transient and may be followed by expansion into a new
phenotypic dimension, creating a new burst of diversification and
subsequent slowdown. Each expansion into a new phenotypic dimension is
associated with an increase in the rate of evolutionary changes of
phenotypes (evolutionary speed), which overall results in a patterns of intermittent burst of evolutionary change and diversification on a background of relative stasis (Fig.~\ref{fig8}).

Putting our results in the context of existing empirical research on the
evolutionary increase of organismal complexity appears to be difficult due to a lack of relevant data. 
Distilling the knowledge about such complexity-expanding transition
from the fossil record is apparently a difficult task, and e.g. determining the timing and
the level of organismal complexity associated with fundamental transitions such as the appearance of predation is still
debated
\cite{bengtson2002origins,kowalewski2002fossil}. Bioinformatics
methods have their own difficulties  caused e.g. by the scarcity of
gene annotations that hinders associating genes with
corresponding phenotypic features. Even at a more basic level,
separating contributions from biotic and abiotic factors to major
evolutionary transitions is a notoriously difficult task (see, for
example, the  review by \cite{voje2015role}), and the role of
adaptation in evolutionary increases of complexity is not resolved 
\cite{lynch2007frailty}.
  
Despite all those difficulties, several established evolutionary facts
can be viewed as supportive of our conclusion. For example, it has
been deduced that the maximum size of organisms has increased mostly
in two discrete steps of approximately equal magnitude
\cite{payne2009two}. Each step required a substantial expansion
in organismal complexity: the first step was associated with the appearance of
the eukaryotic cell, and the second step with eukaryotic
multicellularity. Also, our findings are reminiscent of the notion of rapid adaptive diversification into new
adaptive zone as envisioned in 
\cite{simpson1944}. The appearance of new adaptive zones could be linked to
an expansion to a higher level of phenotypic complexity, which enables
the organisms to function in novel ways and e.g. use novel resources or novel environments.  
More generally, many adaptive radiations \cite{schluter2000} could be viewed from the perspective of increased organismal complexity allowing for expansion into new regions of phenotype space that subsequently cause bouts of diversification. Adaptive radiations may often be perceived to be driven by geological events, such as the colonization of a new and initially empty habitat (e.g. islands or lakes). But migration leading to such colonizations may itself be driven by ecological pressures in the ancestral habitat. Moreover, there are also cases of adaptive radiation that occur in the absence of geological events, such as the radiation in floral diversity in a group of Solanaceae \cite{kostyun2017heterochronic}, which occurred in a period without significant geological changes, and instead was likely caused by competition for pollinators. In general, we think that investigating adaptive radiations as ecologically driven increases in phenotypic complexity and subsequent diversification could be a useful perspective.

Finally, a rather tenuous connection could be seen in a number of well-analyzed examples of convergent evolution \cite
{washburn2016convergent}: multiple appearances of the same potentially complex
trait points to selective forces for their origin, 
and the variation in the timing of the evolution of such traits could indicate that their appearance depends on the presence of the ``right'' ecological scenario.

Naturally, a lot remains to be explored regarding the fascinating question of the evolution of organismal complexity. We
see several immediate possible extensions of our work. First, it
would be desirable to have a more realistic representation of
competition between individuals that live in distinct sets of
dimensions. For example, one should take into account a possible lack
of reciprocity in the competitive effects between high-dimensional and
low-dimensional individuals. Second, other ecological interactions, such
as predation, should be included in future models. In general very little is known about evolution of predator-prey interactions in high-dimensional phenotype spaces.  Finally, it would
be very interesting to investigate evolutionary transitions between different
types of ecological interactions (e.g. from competition to
cooperation) as a particular form of transitions that lead to changes in levels of complexity. Such transitions could also lead to changes in the process of adaptation itself, e.g. due to the appearance of new levels of individuality and multi-level selection.

\section{Acknowledgments}

I.I. was supported by FONDECYT (Chile) grant 1151524. We thank Denis Tverskoy for helpful discussions. 


\end{document}